\documentclass[pre,normal]{revtex4}
\usepackage[dvips]{graphics}
\usepackage{epsfig}
\usepackage{epsfig}
\usepackage{dsfont}
\usepackage{amsmath}

\begin{document}

\title{Derivation of a Fluctuation Theorem from the probabilistic definition of entropy}
\author{W. \surname{Pietsch}\footnote{wpietsch@gmx.de}}
\affiliation{Department for Philosophy of Science, University of Augsburg, Universit\"atsstrasse 10, D-86135 Augsburg, Germany}

\begin{abstract}
It will be shown, how the Boltzmannian ideas on statistical physics can be naturally applied to nonequilibrium thermodynamics. A similar approach for treating nonequilibrium phenomena has been successfully used by Einstein and Smoluchowski treating fluctuations. It will be argued, that due to the reversibility of the microscopic equations, all processes -- also macroscopic ones -- must at least in principle be reversible. Also, a clear conceptual distinction between equilibrium and nonequilibrium states is not possible in the Boltzmannian framework, which is just the reason why these concepts should apply to nonequilibrium. In the present manuscript we derive a Fluctuation Theorem from the equation $S=k_B \ln P$, where $P$ is the probability of a state. The recently discovered Fluctuation Theorems are some of the few exact results valid far from equilibrium. Two assumptions are needed for the derivation: First, the process shall happen on certain time-scales that are large compared with the time, during which the system memorizes its initial conditions. Second, the entropy production rate averaged over all realizations of the process shall be constant during the process. We will finally point out, why a solution to the problem of macroscopic irreversibility invoking causality -- as it was recently suggested in connection with the Fluctuation Theorem -- cannot live up to its task.
\end{abstract}

\maketitle

\pagestyle{headings}

\section{Introduction}
Recent times have shown considerable advance in the field of nonequilibrium thermodynamics, also called thermodynamics of irreversible processes. This important progress -- in a field that is in parts still not well understood -- is closely connected with two related theorems: the Fluctuation Theorem (Evans and Searles, 1993 and 2002) and the Jarzynski Relation (Jarzynski, 1997). Crooks (1999) has shown, that both are closely related, the former being more general than the latter. Let me quote Ruelle (1999) on the importance of the field:
\begin{quote}
``A quantitative approach to situations far from equilibrium has been developed in recent years, 
\ldots\ with such results as the Gallavotti-Cohen fluctuation theorem. This new approach, based on smooth ergodic theory, does not solve all problems. In particular, quantum nonequilibrium is not covered. Nevertheless it appears that, at long last, nonequilibrium statistical mechanics is acquiring the sort of mathematically precise quantitative tools that equilibrium statistical mechanics has possessed and exploited so successfully for more than a century" (p.\ 540). 
\end{quote}
This relatively early (compared with the development in this field) quote refers solely to the so called Steady State Fluctuation Theorem (Gallavotti and Cohen, 1995a and 1995b), by now there is a whole family of closely related Fluctuation Theorems applicable to different physical setups. Fluctuation Theorems determine exactly the relative probability between processes with opposite average entropy production rate: $\overline{\sigma_t}$ and $-\overline{\sigma_t}$. Mainly there are two types: the Transient Fluctuation Theorem is applicable to systems driven away from an initial equilibrium state and the Steady State Fluctuation Theorem is applicable to systems in a non-equilibrium staedy state. In this manuscript we derive a theorem similar to the Fluctuation Theorems from within the framework of conventional statistical physics as developed by Boltzmann, Planck, Einstein, Smoluchowski and others. The conceptual basis of the argument will be, that the interpretation of entropy as probability in principle does not allow a distinction between reversible and irreversible processes as well as equilibrium and nonequilibrium states. Thus, the mentioned theorem will be derived from $S=k_B \ln P$, where $P$ is the probability of a state. It will be shown, how time-averaging allows for the definition of the probability of processes as it allowed for the definition of the probability of states. 

From the philosophical perspective, the context and justification of the present manuscript should be seen in accordance with the view of nonequilibrium thermodynamics as suggested by Jos Uffink (2001):
\begin{quote}
``However, since the Second World War, a lot of work has been done in obtaining extensions of thermodynamics which could be applied to systems out of equilibrium. \ldots\ It is characteristic of this type of work that it is focussed on applications and gives comparatively little attention to the foundations and logical formulation of the theory. \ldots\ The question how the entropy of a non-equilibrium state is to be defined, and the proof that it exists and is unique for all non-equilibrium states, still seem to be largely unexplored" (p.\ 389). 
\end{quote} 
As said, I will rely on the fact, that the Boltzmannian approach does not depend crucially on a distinction between nonequilibrium and equilibrium states, and thus should work equally well for both cases. Within the Boltzmannian framework, there exists a well-defined nonequilibrium entropy, given by the development of the examined system over time. 

This manuscript closely follows the Boltzmannian view on thermodynamics as it has been so fruitfully used by Einstein and Smoluchowski in their treatment of fluctuations, which are of course nonequilibrium phenomena. Due to the conceptual difficulties of the Gibbs approach when treating nonequilibrium, a consequent treatment of fluctuations may in the end be possible only within the Boltzmannian framework. By staying with Boltzmann our conceptual framework does not depend on the ergodicity of a system. However, ergodicity is still a crucial ingredient for explaining the success of the Gibbs approach. In fact, it is still one of the big questions in the foundations of statistical physics, how all the different concepts of entropy fit together. But work is done also on this front (e.g.\ Frigg, 2004). In a first part of this manuscript we will discuss the conceptual foundations both with reference to some historical (but nevertheless relevant) physics-literature and with reference to more recent discussions in the philosophy-of-physics-literature. In the second part we will outline a mathematical framework, that is consistent with the Boltzmann approach and valid also for nonequilibrium states and processes in general. The concepts used owe much to Einstein's early work in statistical physics, e.g. on the theory of critical opalescence (Einstein, 1910) and on quantum mechanics (Einstein, 1905).

\section{Processes in conventional statistical physics}

\subsection{Preliminary note: why Boltzmann should not be underestimated}
It shall shortly be recalled, how the Boltzmannian approach to thermodynamics naturally applies to nonequilibrium. As said, Einstein's and Smoluchowski's theory of fluctuations is a nonequilibrium theory. This extension of the Boltzmann picture is so natural that Einstein repeatedly wondered, why Boltzmann did not draw this conclusion himself. The most famous example of course is Brownian Motion. A little less known is the explanation of critical opalescence given by Smoluchowski (1908) and Einstein (1910). They explain this phenomenon, which is for example responsible for the blue of the sky, with density-fluctuations in the air. Using Boltzmann's statistical entropy for single systems, these density fluctuations of the air correspond to entropy fluctuations. Einstein (1910) presents a very consistent picture of how Boltzmann's ideas apply to nonequilibrium phenomena. 

Let us further examine the case of Brownian Motion. The movement of a Brownian particle is caused by local fluctuations in the velocity distribution around the particle. These fluctuations are only possible within the statistical view of nature. Even more, as will be argued later, they are only possible within the \emph{Boltzmannian} statistical approach. The core of the reasoning for treating Brownian Motion is well grasped by a quote from Smoluchowski (1906) (quoted by Kac, 1986):
\begin{quote}
``N\"ageli thought that he invalidated this theory [that Brownian motion is caused by molecular collisions] by showing that the velocity acquired by a particle of size 0.003 mm in a collision with another particle would only be $2 \cdot 10^{-6}$ mm/s which would be unobservable under a microscope, and claiming further that the shocks coming from all directions would, on the average, cancel. This is the same error in reasoning as that leading to the conclusion that a player in a game of chance (e.g. tossing of a die) would never lose or gain more than a single stake. We know however that good and bad luck do not cancel completely and the longer the game lasts the greater is the average gain or loss'' (p.\ 18).
\end{quote}
In a qualitative explanation for Brownian motion, Smoluchowski assumes, that there is an equal chance for a positive and a negative result. The probability, that within $n$ throws of a dice there are $m$ positive and $n-m$ negative results, is of course
\begin{equation}
\frac{n!}{2^n m! (n-m)!}.
\end{equation}
Then the average deviation (in positive or negative direction) from the value zero, is:
\begin{equation} 
\nu =2 \sum \limits_{m=\frac{n}{2}}^{n} {n \choose m} \frac{2m-n}{2^n}=\frac{n}{2^n} {n \choose \frac{n}{2}},
\end{equation}
which gives for large $n$
\begin{equation}
\nu =\sqrt{\frac{2n}{\pi}}.
\end{equation}
Although this simplified version of Brownian motion is of course not numerically exact, it already shows the correct relation between mean square displacement and time of observation as in the diffusion equation $\langle x^2 \rangle \propto t$. 

The classical thermodynamical view cannot deal with this result, because in it the gas around the Brownian particle should at some point in time have reached equilibrium, where macroscopic parameters shouldn't change (including the velocity of a macroscopically detectable Brownian particle). We also see here, that a consequent thermodynamical viewpoint implying a deterministic Second Law is already incompatible with the molecular structure of matter, in this case the ideal gas. 

Let us now address the curious conceptual defects in the Gibbs approach, when dealing with nonequilibrium. Lebowitz (1999) writes:
\begin{quote}
``However, unlike [the Boltzmann entropy] $S_B$, [the Gibbs entropy] $S_G$ does not change in time even for time-dependent ensembles describing (isolated) systems not in equilibrium. Hence the relevant entropy for understanding the time evolution of macroscopic systems is $S_B$ and not $S_G$" (p.\ S349).
\end{quote}
The problem consists in the fact, that in the Gibbs approach ensembles are used to describe single systems. These ensembles should result from the time-evolution of the system itself (assuming ergodicity). Thus, in the Gibbs approach, essentially time averages are calculated (averaged over infinite times). Thus, there is no room in the Gibbs picture for time evolution. For this very reason, one must resort to Boltzmannian arguments for the treatment of nonequilibrium. Compare also Lavis (2005) for the applicability of the Gibbs approach:
\begin{quote}
``A consistent approach, consonant with our treatment of the Boltzmann approach, is to suppose that the only meaningful probability density function to be used from the Gibbs approach is the time-independent solution of Liouville's equation determined by the dynamics and the physical constraints on the system. A change of physical constraints will lead to an instant discontinuous change in the probability density function and the Gibbs entropy. An uncommon state (like, for example, the case of all the particles being in one end of the box) will have low probability when calculated using [the equilibrium probability density function] $\rho_G(x)$ and low Boltzmann entropy, but the same Gibbs entropy as any other configuration" (p.\ 263).
\end{quote} 
The point of view suggested by Lebowitz and Lavis seems to me the only conceptually sound conclusion drawn from the fact, that in the Gibbsian approach we calculate the entropy of a single system by integrating over an ensemble of systems, including some, that are highly nonequilibrium states from the Boltzmannian point of view.

\subsection{Reversible and irreversible processes}
Generally, processes are divided into reversible and irreversible processes. This distinction has its origin in the 19th century within traditional thermodynamics, i.e.\ in the time before Boltzmann and Planck introduced the probabilistic interpretation of entropy
\begin{equation}
S=k_B \ln P,
\label{eq:S=P}
\end{equation}
where $k_B$ is Boltzmann's constant and $P$ is the probability of a state. The probability $P$ describes the number of microscopic states compatible with the given macroscopic boundary conditions (ensemble-version of probability) or the fraction of time that the system is in the desired macroscopic state (time-version of probability). In the second half of the 19th century Boltzmann suggested a derivation of thermodynamics from mechanistic concepts. The criticism of Poincar\'e, Loschmidt, Zermelo and others mostly relied on the fact that in microscopic physics all processes are reversible while thermodynamics knows irreversible processes. This criticism has contributed in considerable extent to the development of Boltzmann's probabilistic interpretation of entropy, which implies that in principle also all macroscopic processes are reversible (compare also Einstein, 1910, and Smoluchowski, 1915). Only, the process happening in one direction of time is much more probable than its occurrence in the opposite time-direction. In fact, the difference for the probabilities is so large, that most macroscopic processes will be observed only in one direction of time leading to the observed `irreversibility' in the macroscopic world. However, in principle the probabilistic interpretation of entropy is not compatible with a conceptual distinction between reversible and irreversible processes. All processes should be treatable in the same way. Statements as such are essential in the statistical interpretation of thermodynamics. Compare for example Smoluchowski (1916):
\begin{quote}
``A process \emph{appears} irreversible (reversible), when the initial state has a recurrence time which is long (short) compared to the time of observation" (my italics, p.\ 77).
\end{quote}
Uffink (2001) argues, ``that the second law has nothing to do with the arrow of time" (p.\ 305). Since Uffink's discussion concerns thermodynamics, this refers to the thermodynamical, i.e.\ macroscopic, second law. The main merit of his standpoint is, ``that the second law would no longer represent an obstacle to the reconciliation of different theories of physics" (p.\ 94), i.e.\ between thermodynamics and mechanics. Uffink does not seem to favor any of the Boltzmannian approaches to thermodynamics (``But apparently there is another option" (p.\ 388).). Nevertheless, Uffink's task of examining thermodynamics with respect to the distinction between reversible and irreversible processes is of considerable importance also from the Boltzmannian point of view. If a conceptual distinction between reversible and irreversible processes would be possible on a macroscopic scale, that would clearly contrast the Boltzmannian view of the Second Law as stated by Smoluchowski above.

The Boltzmannian explanation for irreversibility is consensus among most physicists and many philosophers of science. Compare for example Ruelle (1993):
\begin{quote}
``The explanation of irreversibility, that we have come up with following Boltzmann, is at the same time simple and quite subtle. It is a probabilistic explanation. It does not imply an irreversibility of the fundamental laws of physics, but there is something special about the initial state of the system, that we are observing: the initial state is \emph{extremely improbable}. \ldots\ I have described the interpretation of irreversibility, that is generally accepted by physicists these days" (my translation, p.\ 119).
\end{quote} 
Or Lebowitz (1993):
\begin{quote}
``Boltzmann's thoughts have withstood the test of time" (p.\ 32).
\end{quote}

The few voices dissenting with the Boltzmannian viewpoint as for example that presented in Prigogine (1999) have thus far not had much impact. Also, the fact that there actually exists a field in physics called the theory of irreversible processes is probably not so much owed to a distrust in the Boltzmannian argument, but stems more from a desire to mathematically reproduce nature at different levels of complexity. The complete reduction of irreversible macroscopic phenomena to the microscopic level, where full reversibility should again show up, is in many cases practically impossible and even undesirable.

\subsection{The probability of states}
For itself Eq.\ (\ref{eq:S=P}) does not make sense without a concept how to determine the probability of a macroscopic state. In general, there have been two different approaches to define the probability of a state. One is the ensemble approach, where the probability is determined by the number of systems that are in the desired states divided by the number of systems that make up the whole ensemble. In the following, we will call this type of probability in short ensemble-probability. This definition is not yet sufficient as such, but strongly depends how one chooses the members of the ensemble. To my knowledge, there are no apparent epistemological reasons, how to choose these members except by relying on the second type of probability.

This type of probability is given by the observation of a single system over the course of a long time-span $\tau$. The probability $P(Z)$ of a certain state $Z$ is then given by the amount of time $\Delta \tau$, that the system is in state $Z$ during $\tau$:
\begin{equation}
P(Z)=\frac{\Delta \tau}{\tau}.
\label{eq:tau}
\end{equation}   
In the following we will shortly call this probability time-probability. Before addressing the difficulties connected with Eq.\ (\ref{eq:tau}), I want to emphasize again that time-probabilities are epistemologically better justified than ensemble-probabilities. Boltzmann, Smoluchowski and Einstein  shared this believe at least for a good part of their life, as they all heavily rely on this concept in their work. As already mentioned, to me the obvious reason to prefer time-probabilities is that it is not clear how to choose the members of an ensemble of systems. It would not make sense to include microscopic states in the ensemble that cannot be reached by the single system considered in an experiment. Thus the ensemble is best chosen by observing which states a single system can take on over the course of time. In other words, this means that ensemble-probabilities should be derived from time-probabilities. 

Another interesting argument in favor of time-probabilities can be found in one of Einstein's most important papers on statistical physics (Einstein, 1910). There, he states that time-probabilities are the most natural approach since they do not have to rely on a microscopic interpretation for the probability of a state. Einstein also emphasizes the connection between time-probabilities and the reversibility of processes. Of course, only if a system will always return arbitrarily close to a state already taken, then the definition (\ref{eq:tau}) will make sense, because there is a well-defined limit as $\tau \rightarrow \infty$. The probability of an area in phase space, which is transversed only a finite number of times, is zero in the limit $\tau \rightarrow \infty$. (Note, that actually the trajectories do not have to be reversible. It suffices to require that a system will always return arbitrarily close to every state once taken on. For example a system consisting of a single particle moving in a fixed direction on a circle is not time-reversible, since a reversal of the trajectory is never observed by definition. However, the time-probabilities of all possible states can well be defined.)

The conceptual problems in the definition of (\ref{eq:tau}) are also noted in Einstein (1910). The main difficulty is, that the more exactly we define the macroscopic state $Z$, the less probable it is. For example if $Z$ is defined through a certain exact value of the macroscopic parameter $\lambda_n=\lambda_n^0$, then if we compare with the cases when $\lambda_n$ is not $\lambda_n^0$, the resulting probability $P(Z)$ will be zero according to (\ref{eq:tau}). 

To resolve this difficulty, we can divide the observation time $\tau$ into intervals of equal length. The probability of a certain state determined by the value $\lambda_n^0$ is then given by the number of intervals, during which $\lambda_n$ assumes the value $\lambda_n^0$ divided by the total number of intervals within $\tau$. (This should be understood just as a general idea, how time-probability could be defined. There may be some technical difficulties like for example normalization, which can however be resolved.) Note, that from the perspective of time-probability this approach is more general than coarse-graining of phase space into cells, as again we do not have to rely on a microscopic theory.

\subsection{Equilibrium and nonequilibrium states}
Just as there is no difference in principle between irreversible processes and reversible processes, there are no conceptual grounds to distinguish between equilibrium and nonequilibrium states. This has been noted for example by Einstein (1917): 
\begin{quote}
``It was clear, that strictly speaking thermodynamic equilibrium does not exist, but rather that every system left to itself will fluctuate irregularly around the state of ideal thermodynamic equilibrium" (my translation, p.\ 107).
\end{quote}
In the framework of statistical physics a macroscopic state is determined by the macroscopic boundary conditions. Once these boundary conditions are fixed, the system should freely make use of its microscopic degrees of freedom (this property called `mixing' is sufficient for the identity of ensemble-probability and time-probability). For example, take the case when a gas is allowed to expand `irreversibly' into empty space. According to conventional dictation the gas is in a nonequilibrium state at first and then relaxes into equilibrium. However, according to the Poincar\'e Recurrence Theorem if we wait long enough, the gas will return arbitrarily close to the supposed nonequilibrium state. In principle the relaxation process will never finish. 
It is generally argued, that the equilibrium state is just the by far most probable state -- this is the late Boltzmannian point of view consistent with Boltzmann's coarse-grained entropy (Boltzmann, 1877). However, this definition of equilibrium depends crucially on which macroscopic parameters one chooses to define equilibrium (we will work this out in more detail below). And then if one looks close enough, one will always detect that the system is subject to more or less tiny fluctuations and thus not in equilibrium. In summary, the system will never be in equilibrium but always in a nonequilibrium state. 

A way out of this dilemma could be the approach of including fluctuations up to a certain amount into the equilibrium state. Van Lith (1999) suggests the following definition:
\begin{quote}
``let us consider a class $\Omega$ of macroscopically relevant quantities and define the system to be in equilibrium from $t=\tau$ onwards iff its distribution $\rho_t$ obeys
\end{quote}
\begin{equation}
\forall F \in \Omega, \; \forall t \geq \tau, \exists c: |\langle F \rangle _{\rho_t}-c| \leq \epsilon_F.
\label{eq:vl}
\end{equation}
\begin{quote}
That is, a system is in equilibrium when the ensemble averages of phase functions in some class $\Omega$ are time-independent, or may fluctuate in time at most within some small, fixed intervals $\epsilon_F$" (p.\ S114).
\end{quote}
As van Lith notes, there rests the problem of determining the $\epsilon_F$. In my view, this point is crucial. Considering the great variety of different systems to which Eq.\ (\ref{eq:vl}) should apply, there may be great difficulties in coming up with a conceptually well founded rule for determining the size of the $\epsilon_F$.  

In a somewhat similar problem runs the late Boltzmann (1877) with the coarse-grained version of his entropy. The six-dimensional phase-space for the single molecule is coarse-grained. The state of the single molecule is determined by the cell of phase-space in which the molecule can be found. The macroscopic state is determined by the number of molecules in each cell. For simplicity of the argument, let us just consider the physical space and assume that the velocity distribution corresponds to the equilibrium distribution. Given a fixed number of particles $N$ in a certain volume $V$, the value of the entropy depends crucially on the size of the cells in space. Let the gas expand from a small part of the volume to the whole volume $V$ (again, the velocity distribution shall not change at all). Now strangely, we will have no change in entropy at all if the cells are chosen small enough such that basically always there is only just one particle in every cell. There may be small fluctuations with finite-size cells -- as sometimes there may be two particles in one cell -- but as the cell size becomes infinitely small we will never have a change in entropy at all. This is grotesquely at odds with the intuitive notion of approach to equilibrium. In the other extreme case, if we let the cell size take the whole volume $V$, then we wouldn't have any change in entropy either. Only for intermediate cell-sizes in physical space, the entropy really changes. We see, how crucial the Boltzmannian definition of equilibrium depends on the cell size. If for example we let the cell size be of size $V$, the system will trivially always be in its most probable state, per definition the equilibrium state. For very small cell size, the system will also always be in equilibrium. Thus, as in the van Lith approach the difficulty consists in determining the size of the $\epsilon_F$, the problem for the Boltzmann equilibrium consists in finding a conceptual foundation for the size of the cells. From the classical point of view, this question has no satisfying answer at all. 

From the quantum-mechanical point of view, the standard argument goes, that the cell size in phase space is determined by the quantum of action. This may be a somewhat satisfying answer, but in any case it is at most only half an answer. The size of the quantum of action does not determine, which `part' of $h$ shall be attributed to the physical space and which one to the velocity-space. In general, the argument presented above still holds, since we can still choose the cell size in physical space as small as possible by adjusting at the same time the cell size in velocity-space. Since the units of space and impulse are not comparable with each other, there is no way to somehow equally distribute $h$. Zeh (1992) disagrees with the whole quantum-mechanical argument for determining the cell size:
\begin{quote}
``the justification of this procedure [i.e.\ the coarse-graining] by the uncertainty relations, and accordingly the choice of the size of these phase space cells as $h^{3N}$ (or $N!h^{3N}$) may be tempting, but would clearly be inconsistent with classical mechanics. The consistent quantum mechanical treatment leads again to the conservation of ensemble entropy (now for ensembles of wave functions). `Quantum cells' of size $h^{3N}$ can be justified only as convenient \emph{units} for measuring the phase space volume in order to obtain the same normalization of entropy as in the classical limit of quantum statistical mechanics, where ensemble entropy vanishes for pure states (which correspond to \emph{one} cell \ldots)" (p.\ 50). 
\end{quote} 
Let us finally quote Lavis (2005) on the subject of getting a proper definition of equilibrium:
\begin{quote}
``At one end there is a well-used concept of equilibrium in thermodynamics and at the other dynamic equilibrium does not exist in measure-preserving reversible dynamic systems" (p.\ 245).
\end{quote}
Lavis suggests to replace the binary concept of equilibrium by a gradual quantity he calls commonness. This seems a natural suggestion, because as we have just argued there will always be a certain arbitrariness in the notion if a system is in equilibrium or not. On the borderline it is hard to imagine any sound reasons, why a certain microscopic state should be counted as an equilibrium state and another one not. Thus, a concept of commonness is certainly much better founded than the binary property of equilibrium. Lavis introduces commonness in the following way:
\begin{quote}
``All references to a system being, or not being, in equilibrium should be replaced by references to the commonness of the state of the system, with this property being given by some suitably-scaled monotonically increasing function of the Boltzmann entropy" (p.\ 257).
\end{quote}
However in my view not much is gained by this proposal. We have introduced a new quantity, that has no real empirical meaning (except of course in the reduction to binary equilibrium), since physicists have thus far been able to interpret nature without it. And since the definition involves a function of entropy, it seems plausible, that we could just work with entropy itself without the need of defining any new entity. On the other hand, if the new concept helps to bridge the gap between the concept of equilibrium according to Gibbs and that according to Boltzmann, this would in my view be reason enough to introduce the new quantity. 

In the continuation of this manuscript I will follow a different line. Keeping in mind the arbitrariness in the definition of equilibrium, maybe the concept itself is not needed at all. I will try to derive a recently discovered theorem from the probabilistic definition of entropy $S=k_B \ln P$, where $P$ is the probability of a state. Here, it is not of importance if $P$ describes the probability for an equilibrium or a nonequilibrium state. 

\subsection{Driven processes and fluctuations}
Driven processes, i.e.\ processes that are determined by a changing external parameter, play an important role in nonequilibrium thermodynamics. Consequently an important question is how driven processes fit into the framework of conventional statistical mechanics. The basic idea here is that every driven process can also be represented as a fluctuation in a system with enough degrees of freedom. The parameters that are changed in a determined way during the driven process must be considered as degrees of freedom for the corresponding fluctuating system. 

Consider for example a gas that is compressed and expanded by a piston. This driven process can be related to a fluctuating process by allowing the volume of the gas fluctuate within the range that the piston covers. Also, since in the driven process the piston performs work on the system while the volume of the gas expands or compresses, the total energy of the gas must be allowed to fluctuate. Of course normally the system will not fluctuate in the way that the piston moves during a determined driven process. However, there should be certain periods in time, where the fluctuation of the system follows the same trajectory in phase space as if the system is subjected to the driven process. These fluctuations can then be evaluated in the framework of conventional statistical physics and the results should be the same as for a theory of driven processes.

In general, $\lambda_1$ \ldots $\lambda_r$ shall be all observable macroscopic variables. We fix a certain amount of these parameters: without loss of generality $\lambda_1$ \ldots $\lambda_f$. The probability of a state, given by certain values of $\lambda_{f+1}$ \ldots $\lambda_r$ will then be determined by the fraction of time during which the systems takes on these values. Now, how can we determine the probability of a state during a driven process? Without loss of generality $\lambda_f$ shall be the external parameter that changes in the process. The probability and thus the entropy of states during a driven process can be determined by looking at a system, where $\lambda_f$ is one of the parameters not fixed.

Note the basic assumption that allows for this interpretation of driven processes: When given a certain amount of freedom, the system will completely cover the whole corresponding phase space in the course of time. The macroscopic degrees of freedom determine the microscopic phase space of the system, in which the system is allowed to move. Notably, there is a connection between which driven processes and which fluctuations are possible. For example, if we allow the system to be in two unconnected areas of phase space, of course it will not fluctuate freely within the whole area, but will only stay in that part where it was at the beginning. For example, these areas could represent two non-overlapping intervals of the parameter $\lambda_n$, which shall change continuously with the microscopic development of the system. It is in accordance with our view, that just as the system will not fluctuate from one area to the other, as well one cannot think of a driven process that connects the two separate areas in phase space.

\subsection{Equilibrium and nonequilibrium processes}
In classical thermodynamics equilibrium processes are defined to be infinitely slow. From the statistical perspective this definition makes no sense, since as discussed above the system will never relax into an equilibrium state, but will always fluctuate more or less far around the most probable state. According to classical thermodynamics, first the boundary conditions are changed and then with constant boundary conditions the system is allowed to `relax'. However, taking seriously the statistical foundation of thermodynamics the system will not relax forever. For example, it will return at later moments arbitrarily close to the state from which it is allowed to relax. Already the use of the expression `relax into' seems to imply an irreversible change which in fact is not happening. For this reason an infinitely slow process is not in any way different from a finite time process, since there is no defined limit for the state of a system as $t \rightarrow \infty$. In principle, there is no difference between equilibrium processes and nonequilibrium processes exactly because the expression `relax into an equilibrium state' is not meaningful in the statistical context. 

In my opinion, the described situation, that there is only one kind of processes and a clear distinction between equilibrium and nonequilibrium processes is not possible as is a sound distinction between reversible and irreversible processes, may well be criticized. For this very reason the discussion about the irreversibility problem is still going on. But one should also keep in mind, that rejecting the situation described in this first part of the manuscript, also means rejecting the Boltzmannian explanation of irreversibility and thus the Boltzmannian interpretation of entropy as probability.

\section{The probability of processes}
\subsection{The time-probability of processes}
Previously, we have discussed the probability of states resulting from the fraction of time, during which the system is in the considered state compared with the full observation time. In a similar manner we can now define the probability of processes consisting of a sequence of states. We can for example determine the relative probability of two processes $A\rightarrow B$ and $C \rightarrow D$ by counting how often process $A \rightarrow B$ is observed during a certain time compared with process $C \rightarrow D$. It is useful but not necessary to assume that both processes should have the same duration $t$. Similar problems are encountered as in the definition of the probability of states. As a process is defined more and more exactly through macroscopic parameters the process will eventually have probability zero. Thus, processes should be determined only up to a given uncertainty. We have for the ratio of probabilities $P$
\begin{equation}
\frac{P(A \rightarrow B,t)}{P(C\rightarrow D,t)}=\frac{N_{A \rightarrow B,t}}{N_{C \rightarrow D,t}},
\end{equation}
where $N$ denotes the number of times the process occurs during the whole observation time. 

\subsection{Derivation of a Fluctuation Theorem}
In the first part of this manuscript it was argued that once the interpretation of entropy as probability is accepted then in principle the distinction between reversible and irreversible processes cannot be made, as well as between equilibrium and nonequilibrium states. This means that the results of so-called nonequilibrium thermodynamics should be attained from the same fundamental concept as the results of equilibrium thermodynamics. Consequently, in the following we will show how a Fluctuation Theorem can be derived from the most basic formula of statistical physics (\ref{eq:S=P}). Only two additional assumptions have to be made. The first comparing the time during which the system loses memory of initial conditions with the time-scales of the considered processes. The second assumption states that the entropy production rate should be constant during an averaged fluctuation.

We can calculate the probability for a fluctuation with help of the following formula:
\begin{equation}
P(A \rightarrow B,t)=P(A \rightarrow Z_1,\Delta t) P(Z_1 \rightarrow Z_2, \Delta t) \ldots P(Z_N \rightarrow B, \Delta t)=P(Z_1)^{\frac{\Delta t}{t}} P(Z_2)^{\frac{\Delta t}{t}} \ldots P(B)^{\frac{\Delta t}{t}}
\label{eq:ass1}
\end{equation}
For the second equality the first assumption introduced above is necessary. The time span $\Delta t$ must be large compared with the time $t_{ml}$ during which the system looses the information about its initial conditions. This is necessary because otherwise, the system could not be in an arbitrary macroscopic state within $\Delta t$. Consider for example the probability $P(Z_1 \rightarrow Z_2, \Delta t)=P(Z_2)^{\frac{\Delta t}{t}}$. If within $\Delta t$ the system could not in principle reach an arbitrary macroscopic state, i.e. if the memory of the initial state $Z_1$ would not be lost, then it would not be consistent to take the probability of the state $P(Z_2)$ as probability for the process $Z_1\rightarrow Z_2$.

A sense for the order of magnitude of $t_{ml}$ can be derived from the typical `relaxation times' of a system, which it takes to attain `equilibrium' starting from a `nonequilibrium state'. Consider the example of a gas expanding into empty space. Usually this expansion is very fast and much faster than a controlled expansion with the use of a moving piston. Thus, just as the assumption seems justified in this particular case, it may be a plausible condition, that however has to be verified in each case. Later on we will discuss qualitatively the limitations that the described condition imposes on the applicability of the Fluctuation Theorem derived here.  
 
Note, that only with $\Delta t$ in the exponent of the probabilities, a consistent definition of the probability of a process is possible. In particular a limiting value can be defined as $\Delta t$ becomes very small (the probabilitites $P$ for the states must be fixed during this limiting process). Also, if we consider a static process, where the system is always in the same state $A$, then the probabilty for the process $P(A\rightarrow A,t)$ is only independent of $\Delta t$ if it is in the exponent. We now employ the basic formula of statistical physics $S=k_B \ln P$. This yields 
\begin{equation}
P(A \rightarrow B,t)=e^{\frac{S(Z_1)+S(Z_2)+S(Z_3)+\ldots+S(Z_N)+S(B)}{k_B}\frac{\Delta t}{t}}.
\end{equation}
We see that this definition for the probability of a process leads to the following result, when we consider a process $A\rightarrow B$ and its time-reversed counter-part $B\rightarrow A$:
\begin{equation}
P(A \rightarrow B,t)=e^{\frac{S(Z_1)+S(Z_2)+\ldots+S(Z_N)+S(B)}{k_B}\frac{\Delta t}{t}}\approx e^{\frac{S(Z_N)+S(Z_{N-1})+\ldots+S(Z_1)+S(A)}{k_B}\frac{\Delta t}{t}}=P(B \rightarrow A,t).
\end{equation}
In the limit $\Delta t \rightarrow 0$ this result is exact. The fact that a process and its time-reversed counter-part both have the same probability is a necessary result for the probabilistic interpretation of entropy. This can be seen when we consider a process that is described by one macroscopic parameter $\lambda_n$ (Ehrenfest and Ehrenfest, 1911, and Smoluchowski, 1916). The most probable value of $\lambda_n$ shall be denoted as $\lambda_n^0$. It is generally assumed in statistical physics that the most probable value is by far more probable than different values of $\lambda_n$. Thus the system will fluctuate around $\lambda_n^0$. Now, when we observe the system over a long enough time $\tau \rightarrow \infty$ of course for every process, where the system fluctuates away from the equilibrium $\lambda_n^0$ by a certain value $\Delta \lambda_n$, there must be a process, where the time-reversed fluctuation takes place. The system will always return to equilibrium. Additionally, it has to be assumed that the underlying microscopic equations are symmetrical in time, so that fluctuations will be similar both towards and away from equilibrium. (E.g.\ without the reversibility of microscopic equations one could imagine, that the required duration for a fluctuation $\Delta \lambda_n$ towards the equilibrium value could in average be different than the duration for a fluctuation $\Delta \lambda_n$ away from the equilibrium value.) 

As is examined by the Fluctuation Theorems (Evans and Searles, 2002), we can now compare two processes (e.g. two fluctuations) both with a duration $t$. Similar to the assumptions for the Transient Fluctuation Theorem, where the initial state is always the equilibrium state, we assume that here the initial macroscopic state for both processes is the same $A$. In one process the entropy shall increase by $S_0$, in the other the entropy shall decrease by $S_0$. We compare these fluctuations $A\rightarrow B$ and $A \rightarrow C$:
\begin{equation}
\frac{P(A \rightarrow B,t)}{P(A \rightarrow C,t)}=\frac{e^{\frac{[S(Z_1)+S(Z_2)+\ldots+S(Z_N)+S(B)]}{k_B}\frac{\Delta t}{t}}}{e^{\frac{[S(Y_1)+S(Y_2)+\ldots+S(Y_N)+S(C)]}{k_B}\frac{\Delta t}{t}}}
\label{eq:fluct-th}
\end{equation}
Now in general we arrive from (\ref{eq:fluct-th}) at a form similar to that of the Fluctuation Theorem only if we assume, that we have
\begin{equation}
S(t+\Delta t)-S(t)=\Delta S,
\label{eq:ass2}
\end{equation}
where $\Delta S$ does not depend on $t$ but only on $\Delta t$, i.e.\ the entropy production rate shall be constant during the whole process (we will later examine, what this means and how this condition can be generalized). Under this condition, we have
\begin{equation}
S(Z_1)+S(Z_2)+S(Z_3)+\ldots+S(Z_N)+S(B)=\frac{N+1}{2}(S(B)+S(Z_1))\approx \frac{t}{\Delta t} \frac{S(B)+S(A)}{2}.
\label{eq:2}
\end{equation} 
Eqs.\ (\ref{eq:fluct-th}) and (\ref{eq:2}) lead to a Fluctuation Theorem, that corresponds in the form to the family of Fluctuation Theorems (Evans and Searles, 2002):
\begin{equation}
\frac{P(\overline{\sigma_t}=\frac{S_0}{t})}{P(\overline{\sigma_t}=-\frac{S_0}{t})}=\frac{P(A \rightarrow B,t)}{P(A \rightarrow C,t)}=\frac{e^{\frac{S(B)+S(A)}{2k_B}}}{e^{\frac{S(C)+S(A)}{2k_B}}}=e^{\frac{S_0}{k_B}}.
\label{eq:FT}
\end{equation}
Here, $\overline{\sigma_t}$ denotes the average entropy production rate of the process. For the last equality we employed $S(B)=S(A)+S_0$ and $S(C)=S(A)-S_0$. 

\subsection{Generalization}
Thus far, we have only derived a Fluctuation Theorem for processes, which exhibit a constant entropy production rate. In general, real processes do not have this property. Consequently we will now show, how the assumption of constant entropy production rate can be generalized. We will see, that the Fluctuation Theorem derived in this manuscript also holds for processes where the entropy production rate is constant only in average over all realizations of the process. This way we can calculate the probabilities for different processes where just the averaged entropy production rate over the whole time $t$ is given. This corresponds to the formulation of the Transient Fluctuation Theorem.

We generalize Eq.\ (\ref{eq:ass1}) in the following way
\begin{equation}
P(A \rightarrow B,t)=P(A \rightarrow Z_1,\Delta t_1) P(Z_1 \rightarrow Z_2, \Delta t_2) \ldots P(Z_N \rightarrow B, \Delta t_{N+1})=P(Z_1)^{\frac{\Delta t_1}{t}} P(Z_2)^{\frac{\Delta t_2}{t}} \ldots P(B)^{\frac{\Delta t_{N+1}}{t}},
\label{eq:ass1-rev}
\end{equation}
where in general the different $\Delta t$ are not equal. The average process is now defined by the geometric mean of all $s$ realizations of the process, where in time $t$ the system undergoes the sequence of states $A$, $Z_1$, \ldots, $Z_N$, $B$:
\begin{equation} 
\nonumber \overline{P(A \rightarrow B,t)}=[P'(A \rightarrow B,t) P''(A \rightarrow B,t) P'''(A \rightarrow B,t)\ldots]^{1/s}=
\end{equation}
\begin{equation}
\nonumber =[P(Z_1)^{\frac{\Delta t_1'}{t}} P(Z_2)^{\frac{\Delta t_2'}{t}} \ldots P(B)^{\frac{\Delta t_{N+1}'}{t}} P(Z_1)^{\frac{\Delta t_1''}{t}} P(Z_2)^{\frac{\Delta t_2''}{t}} \ldots P(B)^{\frac{\Delta t_{N+1}''}{t}} P(Z_1)^{\frac{\Delta t_1'''}{t}} P(Z_2)^{\frac{\Delta t_2'''}{t}} \ldots P(B)^{\frac{\Delta t_{N+1}'''}{t}} \ldots ]^{1/s}=
\end{equation}
\begin{equation}
=P(Z_1)^{\frac{\frac{1}{s}\sum\limits_{1,\ldots,s}\Delta t_1}{t}} P(Z_2)^{\frac{\frac{1}{s}\sum\limits_{1,\ldots,s}\Delta t_2}{t}} \ldots P(B)^{\frac{\frac{1}{s}\sum\limits_{1,\ldots,s}\Delta t_{N+1}}{t}}
\end{equation}
We see that the Fluctuation Theorem (\ref{eq:FT}) holds under the following condition: If without loss of generality the states $A$, $Z_1$, \ldots, $Z_N$, $B$ are chosen such that their entropies differ by a constant value $\Delta S$, it follows directly
\begin{equation}
\frac{1}{s}\sum\limits_{1,\ldots,s}\Delta t_1=\frac{1}{s}\sum\limits_{1,\ldots,s}\Delta t_2=\ldots=\frac{1}{s}\sum\limits_{1,\ldots,s}\Delta t_{N+1}=\Delta t.
\end{equation}
This statement is the final formulation of the second assumption. In other words, the entropy production for a process must be constant in average over all realizations of the process. 

\subsection{Applicability of the Fluctuation Theorem (\ref{eq:FT})}
In the following we want to briefly discuss the two assumptions made in the course of the derivation of the Fluctuation Theorem (\ref{eq:FT}). 

As noted before, Eq.\ (\ref{eq:ass1}) is valid, if we assume that the rate at which the system looses information about its initial conditions is much greater than the rate of the considered processes (first assumption). This condition imposes a limitation on the validity of the Fluctuation Theorem (\ref{eq:FT}). Accordingly, processes or fluctuations happening too fast should not be described by the Fluctuation Theorem (\ref{eq:FT}). In comparison, the Fluctuation Theorems (Evans and Searles, 2002) at first do not seem to set any restrictions on the rate of considered fluctuations, i.e.\ on the time in which a given fluctuation may occur. Compare for example the Transient Fluctuation Theorem as given by Evans and Searles (2002, p.\ 1532):
\begin{equation}
\frac{p(\overline{\Sigma_t}=A)}{p(\overline{\Sigma_t}=-A)}=e^{At}\; \forall\; t.
\label{eq:evans}
\end{equation}
Here, $\overline{\Sigma_t}$ is the entropy production rate and $p(\overline{\Sigma_t}=A)$ denotes the probability that the value of $\overline{\Sigma_t}$ lies within the range $A$ to $A+dA$. As no restrictions are made for the size of $A$ by Eq.\ (\ref{eq:evans}), no restrictions are made on the rate of the considered fluctuations. However, the microscopic equations do impose restrictions on the size of $A$. If the energy of the considered system has an upper boundary, there exists also an upper boundary for the maximum velocity of the particles which make up the system. This boundary on the velocity means that certain areas of the phase space cannot be connected within an arbitrarily small time-span. Consequently, there results also a maximum on the entropy production rate in the system, which is not determined by (\ref{eq:evans}). 

The second assumption states that the averaged entropy production rate is always constant if we consider a process with the sequence of macroscopic states $A$, $Z_1$, \ldots,$Z_N$, $B$. Eq.\ (\ref{eq:ass2}) translates to
\begin{equation}
\frac{P(t+\Delta t)}{P(t)}=\textnormal{const}
\label{eq:pt}
\end{equation}
independent of $t$. As we average over all possible realizations of the considered process, of course there can be only one average entropy production rate for every instant of the process. This does not determine if the entropy production rate will change over the course of the process or not. However, condition (\ref{eq:pt}) seems quite natural: It states that the average time needed for a fluctuation from a state $Z_n$ to another state $Z_{n+1}$ depends only on the ratio of the number of microscopic states consistent with each macroscopic state. 

Of course, both assumptions must be verified for each system examined. The fact, that this task is not at all easy does not alter our main point concerning the role of the Fluctuation Theorem within the framework of statistical physics.

\section{Addendum: on the concept of causality required by reversibility}
We will in the following shortly develop a concept of causality that is consistent with the reversibility of microscopic equations, because by Evans and Searles (1996) causality is claimed as a solution to the problem of macroscopic irreversibility. We will show, why this proposal cannot live up to its task. The following section is motivated by a statement in \textit{The physical basis of the direction of time} of Zeh (1992):
\begin{quote}
``The laws of nature, thus refined to their purely dynamical sense, describe the time dependence of physical states, $z(t)$, in a general form -- usually by means of differential equations. They are called \emph{deterministic} if they uniquely determine the state at time $t$ from that at an earlier or later time (and possibly its time derivative), that is, from an appropriate initial or final condition. This symmetric causal structure of dynamical determinism is stronger than the traditional concept of \emph{causality}" (p.\ 1).
\end{quote}
Obviously, Zeh argues for a symmetric, i.e.\ reversible, causal structure of the microscopic equations, that is of course not compatible with the intuitive notion of irreversible causality. Thus, it is straight-forward to conclude that the problem of macroscopic irreversibility cannot be solved by invoking an `irreversible' concept of causality. In {\it Physics and Chance} Sklar (1993) argues in a similar way:
\begin{quote}
``It is certainly quite conceivable that with sufficient insight we could construct an argument that starts with the standard temporal asymmetry of temporarily isolated systems in a time parallel process of entropic increase in one time direction and not the other, and end with a plausible generation of the conditions in the natural world that ground our intuition of causation going from past to future or of records being only of the past and not the future. Perhaps, we could even explain the one-way efficacy in time of the process we call memory. It remains to be seen if this is so. This final stage of the Boltzmann thesis is neither proven nor disproven at the moment" (p.\ 404).
\end{quote}
The quote of Sklar is revealing in two aspects. First, he argues that the asymmetric notions of causality could very well be \emph{derivable} from reversible microscopic equations for a temporarily isolated system. Second, he argues that this program should be worked out within the Boltzman approach. 

The most important conclusion we can draw from both quotes, is that the irreversibility problem cannot be solved starting from an intuitive concept of asymmetric causality which is not somehow expressed in the microscopic equations. Basically two options are left for the role causality could play in the irreversibility context. First, causality could be important on the axiomatic level of physical theories. Then, the fundamental equations would have to be changed in order to represent irreversible causality. Second, the microscopic equations are correct. Then `irreversible' causality is only a derived concept, a name for a property of the physical equations. Because it is a derived concept it cannot provide an alternative for the Boltzmannian solution to macroscopic irreversibility. It's only a new name for it. Since Evans and Searles (2002) do not claim to change the fundamental equations to express irreversibility (p.\ 1529, this was for example the approach taken by Prigogine, 1999), their `solution' must be option two, i.e.\ it must correspond to the Boltzmannian picture. 

This said, we will in the following work out in more detail, what concept of causality is required by reversible microscopic equations. In accordance with the quote from Zeh, given above, this reversible causality strongly contradicts the notion of causality, as it is presented by Evans and Searles (2002): \begin{quote}
``The future state of the system is computed solely from the probabilities of events in the past. This is called the Axiom of \emph{Causality}. 

It is logically possible to compute the probability of occurrence of present states from the probabilities of future events, but this seems totally unnatural. {\it Will the electric light be on now, because at some time in the (near) future, we will throw a switch which applies the necessary voltage} [my italics]? A major problem with this approach is that at any given instant, the future states are generally not known! In spite of these philosophical and practical difficulties, we will explore the logical consequences of the (unphysical) Axiom of Anticausality" (p.\ 1564). 
\end{quote}
We will first examine causality in Newtonian physics and find that the Newtonian Laws do not allow for a conceptual distinction between cause and effect, in particular a temporal asymmetry cannot be established. The third Newtonian axiom `actio=reactio' states the equivalence of the acting and the reacting force. In particular this means that one cannot distinguish in principle, which is the acting and which the reacting force, i.e.\ cause and effect cannot be distinguished in principle. Consider two massive particles A and B acting upon each other. The properties of particle A, e.g.\ its mass, are responsible for the motion of particle B and vice versa. It cannot be determined `which acceleration of which particle comes first'. If there are no forces at all, then there are no accelerations and the state of the system stays the same, which is also a time-symmetric setup. There is no distinction between cause and effect through time-order, as cause and effect (accelerations of the two particles) are happening simultaneously (velocity of interaction is infinite). This can be seen very clearly in Newton's law of gravity
\begin{equation}
F=G\frac{m_A m_B}{r^2}.
\end{equation}
Here, the physical origin for the fact, that cause and effect are interchangeable, is the identity of passive and active mass for every particle. This corresponds to the third Newtonian axiom. Thus, in Newton's theory there is only \emph{inter-}action. Cause and effect always have the same time-coordinate and are exchangeable.

Accepting the conceptual equality of cause and effect directly leads to the reversibility of all microscopic equations and vice versa.
In fact, the Newtonian laws are only time-reversible since they rely on a concept of interaction and not time-ordered causation. This holds for the Hamiltonian formulation of mechanics as well as the equivalent Newtonian formulation. 

In Maxwellian electrodynamics, the situation might at first seem different, since in the framework of this theory signals can travel at most with the speed of light. Thus, at first sight there seems to be a fixed time-order of causes and effects. Still, Maxwell Equations are known to be invariant under time reversal and simultaneous reversal of the magnetic field. It is the existence of both advanced and retarded solutions that makes Maxwellian electrodynamics invariant under time reversal. When advanced solutions are dismissed on the basis of `causality', as is done in many text-books on electrodynamics, already the reversibility of the theory is given up. So, the reversibility of Maxwellian electrodynamics depends on the acceptance of advanced solutions, in other words solutions where in a conventional dictation cause and effect would be exchanged. Only under this premise is the theory time-reversible because only then it is not possible to distinguish between cause and effect because every process can happen in both directions: For every process with cause A and effect B one must accept the possibility of a process with cause B and effect A. In addition we do not make any assumptions about the probabilities of those processes (which in fact is not possible in the framework of Maxwellian electrodynamics). 

If in the following we speak of time-ordered causality, we do not refer to the reversible causality just described for the case of electrodynamics but imply that cause and effect are not conceptually equivalent. In a strong version of time-ordered causality at least for some processes with cause A and effect B there does not exist a corresponding process with cause B and effect A. In a weaker version of time-ordered causality, for every process with cause A and effect B there exists a corresponding process with cause B and effect A, but the probabilities of both processes differ. The weaker version is essentially the solution to the irreversibility problem as given by Boltzmann. This solution is possible because the microscopic equations do not make any statements about the probabilities of processes. They only state, if processes are allowed in principle or not.

It is tautological to say that time-ordered causality can solve the problem of irreversibility. The microscopic equations do not exhibit time-ordered causality. Once we admit a clear distinction between events of cause and events of effect, then of course physics is not reversible by definition. But this distinction should also be represented in the fundamental equations.

\section{Conclusion}
The main point of this text was to show, that the probabilistic foundations of statistical physics as developed by Boltzmann, Planck, Einstein, Smoluchowski and others are of a very general nature and should if correct incorporate both equilibrium and nonequilibrium thermodynamics. Identifying entropy with probability leads to the possibility of fluctuations and renders all processes reversible. The system will always fluctuate around the most probable state and thus the system will never be exactly in an equilibrium state (corresponding to the most probable state). Conceptually it is not possible to adequately distinguish between equilibrium and nonequilibrium states. Consequently, we have set out in this work to derive one of the recent results of nonequilibrium thermodynamics starting from the foundations of general statistical physics. With two additional assumptions a Fluctuation Theorem can be derived only from $S=k_B \ln P$. Of course, the argument followed in this manuscript does not render the results of nonequilibrium thermodynamics (or thermodynamics of irreversible processes) redundant. It only suggests how they might fit into the framework of general statistical physics. Accepting the probabilistic interpretation of entropy, a statistical physics of processes should be possible within the general framework of statistical physics.

\end{document}